\documentclass[prc,a4paper,preprint,showpacs,byrevtex]{revtex4}

\usepackage{graphicx}
\usepackage{dcolumn}
\usepackage{amsmath}
\usepackage{array}
\usepackage{bm}
\usepackage{textcomp}
\usepackage{amssymb}

\begin{document}

\title{Meson and glueball spectra with the relativistic flux tube model}

\author{Fabien \surname{Buisseret}}
\thanks{FNRS Research Fellow}
\email[E-mail: ]{fabien.buisseret@umh.ac.be}
\affiliation{Groupe de Physique Nucl\'{e}aire Th\'{e}orique,
Universit\'{e} de Mons-Hainaut,
Acad\'{e}mie universitaire Wallonie-Bruxelles,
Place du Parc 20, BE-7000 Mons, Belgium}

\date{\today}

\begin{abstract}
The mass spectra of heavy and light mesons is computed within the framework of the relativistic flux tube model. A good agreement with the experimental data is obtained provided that the flux tube contributions, including retardation and spin-orbit effects, are supplemented by a one-gluon-exchange potential, a quark self-energy term and instanton-induced interactions. No arbitrary constant is needed to fit the absolute scale of the mass spectra, and the different parameters are fitted on lattice QCD in order to strongly restrict the arbitrariness of our model. The relevance of the present approach is discussed in the case of glueballs, and the glueball spectrum we compute is compared to the lattice QCD one. Finally, we make connections between the results of our model and the nature of some newly discovered experimental states such as the $f_0(1810)$, $X(3940)$, $Y(3940)$, \textit{etc.}
\end{abstract}




\maketitle

\section{Introduction}
Potential quark models have been proved to successfully reproduce the experimental meson and baryon mass spectra for nearly thirty years~\cite{old,old2,old3,instanton,brau02,russe1,russe2}. Since the pioneering works on this subject~\cite{old,old2}, they have been a matter of constant interest. In particular, new hadrons are continuously being discovered, and potential models offer an intuitive and efficient way to understand the physical properties of these new experimental states. Apart from many other relevant effective approaches of QCD~\cite{effec}, lattice QCD has recently emerged as a powerful method to deal with the full QCD theory (see Ref.~\cite{lint} for an introduction). Interestingly, some basic features of potential models in the heavy meson sector have been confirmed by lattice QCD calculations. The potential energy between a static quark and antiquark has indeed been shown to be roughly compatible with a potential of the Cornell form $ar-\kappa/r+C$~\cite{balirep,barc891}, which is widely used in potential models~\cite{Lucha}. The total interaction is then seen as the sum of a long-range part which encodes the confinement and a Coulomb term arising from one-gluon-exchange process between the quark and the antiquark. Finally, $C$ is an arbitrary negative constant used to fit the absolute scale of a given mass spectrum. 

The structure of light meson spectra can also be understood by using potential models~\cite{old2}. In particular, a semirelativistic quark kinetic term of the form $\sqrt{\vec p^{\, 2}+m^2}$ allows to deal with light, and even massless, particles. As suggested by the background perturbation theory~\cite{Simo00}, the relativistic, spin-dependent, corrections, can then be developed in powers of $1/\mu^2$, $\mu$ being defined by $\left\langle \sqrt{\vec p^{\, 2}+m^2}\right\rangle$. In this expression, the average values are typically computed with the eigenstates of a spinless Salpeter Hamiltonian with the Cornell potential. Such a framework leads to interesting results concerning light mesons~\cite{russe2,bada02}, and reduces to the usual formalism for heavy mesons, since $\mu\approx m$ at large quark mass. Developing the relativistic corrections in powers of $1/\mu^2$ is rather natural: It is actually nothing but developing them in terms of the quark energies rather than in terms of their masses. Such a procedure has already been successfully applied to light mesons in many previous potential models of mesons and baryons~\cite{old2,Lucha}.  

A more recent approach, called the relativistic flux tube model, is an effective meson model obtained from the full QCD theory \cite{QCDstring2,tf_1,equiv,ch_aux_bib}. It relies on the assumption that the quark and the antiquark in a meson are linked by a straight color flux tube, actually a Nambu-Goto string, carrying both energy and angular momentum. This string, or flux tube, is responsible for the confinement. It is worth noting that lattice QCD simulations show that the chromoelectric field between a static quark-antiquark pair is roughly constant on a straight line joining these two particles \cite{Koma}, validating the physical picture of the relativistic flux tube model. Like potential models, it produces linear Regge trajectories in the ultrarelativistic limit, that is a linear link between the square masses of light mesons and their angular momentum, and it reduces to the usual Schr\"{o}dinger equation with a linear potential in the heavy quark limit. We made in previous works several attempts to include in the relativistic flux tube model physical mechanisms that were usually neglected: Quark self-energy~\cite{buis05a}, retardation effects~\cite{buis05b,buis07a}, and spin interactions~\cite{buis07_so}. Apart from these corrections, that are all due to the flux tube itself, it is also necessary to include short-range potentials such as the Fermi-Breit interaction, which comes from the one-gluon-exchange diagram between a quark and an antiquark, in order to get a satisfactory description of mesons with the relativistic flux tube model~\cite{tf_Semay,oge}. 

Our aim in the present paper is to put together all these physical contributions, which were up to now only separately studied. By doing this, we will build a generalized relativistic flux tube model leading to mass spectra in quantitative agreement with the experimental data, and to relevant suggestions concerning the nature of some recently discovered states such as the $f_0(1810)$, $X(3940)$, $Y(3940)$, \dots This model is presented in Sec.~\ref{model}. Then, we apply it to compute heavy and light meson mass spectra in Secs.~\ref{hmes} and \ref{lmes}, where we systematically compare our data to the available experimental states. The application of potential models to glueballs -- bound states of gluons -- has already been done in the past~\cite{bar,corn,hou,Brau04,new,abre}, in good agreement with lattice QCD results. However, the validity of such models to describe glueballs in still controversial. We discuss in Sec.~\ref{gb} the possible extension of our relativistic flux tube model to the case of a bound state of two gluons. Finally, we study the mixing between scalar mesons and glueball in Sec.~\ref{mixi}, and draw some conclusions in Sec.~\ref{conc}.

\section{The model}\label{model}
\subsection{Hamiltonian}
Starting from the full QCD theory, a Lagrangian for a system of two confined spinless color sources can be derived: The corresponding model has been called the rotating string model \cite{QCDstring2}. In this approach, originally introduced as an effective meson model, the quark and the antiquark are linked by a Nambu-Goto string, or flux tube. This object, characterized by its energy density $a$, simulates the exchange of gluons responsible for the long-range part of the interaction, that is the confinement. The rotating string model is completely equivalent to another phenomenological description of mesons, the relativistic flux tube model~\cite{tf_1}, once the auxiliary fields (or einbein fields) appearing in the rotating string model are properly eliminated \cite{equiv,ch_aux_bib}. Although it can be numerically achieved, the resolution of the nonlinear coupled equations of the relativistic flux tube model is a difficult problem \cite{buis041}. More convenient expressions can be obtained by using an accurate approximation of this approach, that we previously called the perturbative flux tube model~\cite{bada02,buis05a}. The present section is devoted to the presentation of the particular Hamiltonian underlying this model. 

The complexity of the relativistic flux tube equations comes from the fact that the contribution of the flux tube does not reduce to a simple static potential: This object is dynamical, and carries orbital angular momentum as well as energy. When the orbital angular momentum of the meson, denoted as $\ell$, is equal to zero, the two-body relativistic flux tube model becomes a spinless Salpeter Hamiltonian with a linear confining potential $ar$. Such a Hamiltonian has been intensively used in the literature to study the properties of hadrons (see a review in Ref.~\cite{Lucha}). As $a$ is the energy density of the flux tube and $r$ is the quark-antiquark separation, it is readily observed that the linear potential is actually the total energy of a static straight string whose length is $r$. The dynamical contribution of the flux tube is actually small enough to be treated as a perturbation of the spinless Salpeter Hamiltonian with potential $ar$, and one then obtains the so-called perturbative flux tube model \cite{bada02,buis05a,new}. If $\ell$ is not too large (typically $\ell<6$), this approach reproduces the ``exact" flux tube mass spectrum with an accuracy better than $5\%$~\cite{buis05a}. 

For a bound state of two particles of the same current mass $m$, that is the case we restrict to in this paper, the perturbative flux tube model is defined by the following spinless Salpeter Hamiltonian
\begin{equation}\label{h0pftm}
H=2\sqrt{\vec p^{\, 2}+m^2}+ar,	
\end{equation}
completed by a perturbation which encodes the dynamical contribution of the string through its angular momentum \cite{bada02,buis05a}
\begin{equation}\label{dh1pftm}
	\Delta H_{str}=-\frac{a\ell(\ell+1)}{\mu\, r(6\mu+ar)}.
\end{equation}
The quantity $\mu$ appearing in the above equation can be seen as a dynamical quark mass given by
\begin{equation}\label{mupftm}
	\mu=\left\langle\sqrt{\vec p^{\, 2}+m^2} \right\rangle,
\end{equation}
in which the average value is computed with an eigenstate of the Hamiltonian~(\ref{h0pftm}). The dynamical mass is then state
dependent. The contribution $\Delta M_{str}$ to the total mass can be computed with a good accuracy by the following approximation~\cite{bada02,buis05a}
\begin{equation}
	\Delta M_{str}=-\frac{a\, \ell(\ell+1)}{\mu\, (6\mu+a\left\langle r\right\rangle)}\left\langle \frac{1}{r}\right\rangle.
\end{equation}

Equations~(\ref{h0pftm}) and (\ref{dh1pftm}) define the original perturbative flux tube model, as studied in Refs.~\cite{bada02,buis05a}. However, important improvements can be brought by dropping the various simplifying assumptions underlying this approach. Three hypothesis are actually used to establish the equations of the rotating string model, and equivalently of the relativistic flux tube model~\cite{ch_aux_bib}. The first one is the assumption that the flux tube is a straight line linking the quark and the antiquark. As we already pointed out, such an ansatz is in agreement with lattice QCD simulations \cite{Koma}. Moreover, other calculations within the framework of effective models show that the possible deviations of the string from the straight line are negligible~\cite{Alle99,buis07a} in mesons. 

It is worth mentioning that we only focus on usual mesons in the present work. The flux tube model can actually succesfully describe hybrid mesons as well, i.e. mesons in which the gluonic field is in an excited state. But, in the case of hybrid mesons, the linear flux tube picture does not hold anymore: An excitation of the gluonic field deeply changes the configuration of the system. A first possibility of modelizing it within the flux tube model is to treat the fluctuations of the flux tube in a quantized way. Then, the different energy levels of the flux tube define configurations corresponding to hybrid mesons~\cite{Allen:1998wp}. A second way is to assume that the excitation of the gluonic field creates a constituent gluon~\cite{horn}. A hybrid meson is then seen as a quark-antiquark-gluon bound state in which the gluon is linked to the quark and to the antiquark by two straight strings. In every case, we can conclude that a -- weakly deformed or not -- straight line configuration of the flux tube is only valid for usual mesons, when the gluonic field is in its ground state. This is the case that we will treat in the latter.

The second simplification which is made to obtain the relativistic flux tube model is the neglect of the retardation effects. It can be shown that these effects can be computed in perturbation within the framework of the rotating string model~\cite{buis05b}, by allowing the quark and the antiquark to have a different temporal coordinate. Retardation effects bring a contribution to the total mass which reads
\begin{subequations}\label{retar}
\begin{equation}\label{dh2pftm}
	\Delta M_{ret}=-\frac{\beta}{2a_3}, 
\end{equation}
with
\begin{equation}
	\beta^2=\left\langle \frac{a^2}{12}+\frac{a p^2_r}{\mu\, r}+\frac{a\, \mu}{2\, r}+\left[\frac{a^2}{90}-\frac{a p^2_r}{20\, \mu r}-\frac{a\, \mu}{12\, r}\right]y^2\right\rangle,
\end{equation}
\begin{equation}
	a_3=\left\langle \frac{\mu}{2}+\frac{a\, r}{12}+\frac{a\, r\, y^2}{40}\right\rangle,
\end{equation}
\begin{equation}
	y=\frac{\sqrt{\ell(\ell+1)}}{r(\mu+ar/6)}.
\end{equation}
\end{subequations}
In these last formulas, $p_r$ is the quark radial momentum, while $y$ is its transverse speed. The various symmetrizations on the noncommuting operators have not been written in order to clarify the notations. We refer the reader to Ref.~\cite{buis05b} for a detailed study of the retardation effects, including the computation of the relations~(\ref{retar}). However, for our purpose, it is 
interesting to notice that retardation brings a negative contribution to the total mass. The physical content of Eq.~(\ref{dh2pftm}) appears more clearly in two limits:
\begin{equation}\label{retapp}
	\left.\Delta M_{ret}\right|_{m=0}\approx -\frac{3a}{8\mu},\quad \left.\Delta M_{ret}\right|_{m\rightarrow\infty}\propto\left(\frac{a^2}{m}\right)^{1/3}.
\end{equation}
As expected, the retardation effects vanish for heavy quarks, where the typical speed scale is very small with respect to the speed of light. On the contrary, they are maximal for light particles. An interesting feature of $\left.\Delta M_{ret}\right|_{m=0}$ is that it preserves the Regge trajectories~\cite{buis05b}. 

The third simplification is that the rotating string model neglects the spin of the quark and the antiquark. An attempt to exactly include the spin degrees of freedom in the relativistic flux tube model has been made in Ref.~\cite{olss04}, but it leads to very complicated equations, that, to our knowledge, have not been solved yet. However, the spin contribution can be described in perturbation by a spin-orbit coupling between the angular momentum of the flux tube and the spin of the confined particles. In a two-body system where the constituent particles have the same mass, this term reads~\cite{buis07_so}
\begin{equation}\label{dh3pftm}
	\Delta H_{so}=-\frac{a\vec{L}\cdot\vec{S}}{2\mu^2r},\quad \Delta M_{so}=-\frac{a}{2\mu^2}\left\langle \frac{\vec{L}\cdot\vec{S}}{r}\right\rangle,
\end{equation}
where $\vec S=\vec S_1+\vec S_2$ is the total spin of the system, and where $\vec L$ is the relative orbital angular momentum in center of mass frame. This formula agrees with the results of background perturbation theory~\cite{Simo00}, and with those obtained with the Wilson loop technique~\cite{barc891} in the limit of heavy quarks. Such a spin-orbit coupling can actually be thought as a Thomas precession of the spin of the confined particles in the color field. 

Apart from these flux tube corrections, it has recently been shown that the quark self-energy contribution, which is created by the color magnetic moment of the quark propagating through the vacuum background field, adds a negative constant to the hadron masses \cite{Sim1}. In the case of a meson made of two quarks of the same mass, the quark self-energy contribution to the meson mass is given by~\cite{Sim1,buis05a}
\begin{equation}
	\Delta M_{qse}=-\frac{f\, a\, \eta(m/\delta)}{\pi\mu} .
\end{equation}
The $\eta$-function is such that $\eta(0) = 1$, and its value decreases monotonically toward $0$ with increasing quark
mass (its explicit form can be found in Ref.~\cite{Sim1}). $\delta$ is the inverse of the gluonic correlation length,
and its value is estimated at about $1.0-1.3$ GeV by lattice QCD computations~\cite{DiGia1,DiGia2}. As the quark self-energy contribution varies very little with this parameter (less than $1\%$), its value can be fixed at $1$ GeV~\cite{buis05a}. The factor $f$ has been computed by lattice calculations. First quenched calculations gave $f = 4$ \cite{DiGia2}. A more recent unquenched work \cite{DiGia1} gives $f = 3$, the value that we choose in this work. 

Finally, a significant contribution to hadron masses is also given by the one-gluon-exchange mechanism between color sources. These are short-range interactions, which are consequently not encoded in the relativistic flux tube model. The one-gluon-exchange potential for mesons is equal to the well-known Fermi-Breit interaction, that is~\cite[p. 239]{Ynd}, \cite{oge,Simo00}
\begin{eqnarray}\label{sqqpot}
    V_{oge}(\vec r\, )&=&-\frac{4\alpha_S}{3}\left\{U-\frac{1}{\mu^2}\left(\frac{1}{4}-\frac{\vec S^{\, 2}}{3}\right)\Delta U+\frac{3}{2\mu^2}\vec L\cdot\vec S \ \frac{U'}{r}\right.\nonumber\\
    &&\left.+\frac{1}{6\mu^2}\left(U''-\frac{U'}{r}\right)\left[\vec S^{\, 2} -3\frac{(\vec S\cdot\vec r)^2}{r^2}\right]\right\},
\end{eqnarray}
where the Darwin term has been neglected. $\alpha_S$ is a phenomenological strong coupling constant, and $U(r)$ is the gluon propagator in position space. For massless exchanged gluons, one simply has $U(r)=1/r$, but Eq.~(\ref{sqqpot}) remains valid for any form of the gluon propagator~\cite{bla}. 

By putting together all these ingredients, a mass spectrum can be numerically computed. The numerical method that we use in this work is the Lagrange-mesh method, a remarkably simple and accurate method to solve nonrelativistic as well as semirelativistic eigenequations \cite{sem01}. The procedure is the following. Firstly, we compute the eigenvalues $M_0$ of the Hamiltonian
\begin{equation}\label{h00pftm}
	H_0=2\sqrt{\vec p^{\, 2}+m^2}+ar- \frac{4}{3}\, \alpha_S U(r),
\end{equation}
which is basically a spinless Salpeter Hamiltonian with a Cornell potential, encoding the dominant contributions of our model. The $ar$ term comes from confinement, while the $-(4/3)\alpha_S U(r)$ potential is the lowest order contribution of the one-gluon-exchange potential~(\ref{sqqpot}). Secondly, the dynamical mass $\mu$ can be computed thanks to the relation~(\ref{mupftm}). Thirdly, the various contributions we presented here can be added in perturbation to get the total mass $M$, defined by
\begin{equation}\label{masss}
	M=M_0+\Delta M_{str}+\Delta M_{ret}+\Delta M_{so}+\Delta M_{qse}+\left\langle V_{oge}(\vec r)+\frac{4}{3} \alpha_S U(r)\right\rangle.
\end{equation}
Clearly, $M$ depends on the radial and orbital quantum numbers $n$ and $\ell$, but also on the spin degrees of freedom. The spectroscopic notation $(n+1)^{2S+1}\ell_J$ will be used in order to unambiguously refer to a particular state. We recall that the parity and the charge conjugation of a meson are given by $P=(-1)^{\ell+1}$ and $C=(-1)^{\ell+S}$. 

Let us stress two interesting advantages of the flux tube model we presented in this section. Firstly, it clearly appears that the mass formula we obtain is valid for any value of $m$. The reason is that the different corrections to the dominant Cornell potential are developed in powers of $1/\mu$ rather than $1/m$. For heavy quarks, $\mu\approx m$, and we recover usual relativistic corrections of order $1/m^2$. However, there is a hope that this model can be generalized for light quarks since, even if $m=0$, $\mu$ remains approximately equal to $0.3$ GeV~\cite{new}. It is worth mentioning that older nonrelativistic quark models phenomenologically used a dynamical mass around this value of $\mu$, with a satisfactory agreement with experiment~\cite{old}. Secondly, we do not need to add a negative constant to the Cornell potential in order to fit the absolute mass scale of the spectrum, as it is the case in most of the potential models~\cite{Lucha,old,old2,instanton}. Formally, this constant can be thought to be replaced by the recently found quark self-energy and retardation terms, which are state-dependent.

\subsection{Effective quark size}
Within the framework of a potential QCD model, it is natural to assume that a quark is not a pure point-like particle,
but an effective degree of freedom that is dressed by a gluon and quark-antiquark pair cloud. Such an hypothesis leads to very good results in the meson \cite{instanton} and baryon \cite{brau02} sectors, and can also be applied to glueballs~\cite{Brau04,Math}. As in these two last references, we assume a Yukawa color charge density for the confined particles
\begin{equation}
	\rho(\vec u)=\frac{1}{4\pi\gamma^2}\frac{{\rm e}^{-u/\gamma}}{u},
\end{equation}
where $\gamma$ is the effective size. It is readily checked that $\rho(\vec u)$ reduce to $\delta^3(\vec u)$ when $\gamma$ tends to zero. The interactions between the two confined particles are then modified by this density, a bare potential $V(r)$ being transformed into a dressed potential $\tilde V(r)$. This potential is obtained by a double convolution over the densities of each interacting particle and the bare potential. It can be shown that, when both confined particles have an equal effective size $\gamma$, this procedure is equivalent to the following calculation~\cite{silv,Brau04} 
\begin{equation}
\tilde V(\vec r)=\int d^3r'\, V(\vec r\ ')\, \Gamma(\vec r-\vec r\ '),\ {\rm with}\ \Gamma(\vec u)=\frac{{\rm e}^{-u/\gamma}}{8\pi\gamma^3}.
\end{equation}
This is actually our case, since we only deal with bound states of two particles of the same mass. So, we can assume that their effective sizes are identical. Other color-charge densities could be used, a Gaussian one for instance \cite{instanton}. We have nevertheless strong indications
that such a change cannot noticeably modify the results \cite{brauth}; moreover, all convolutions are analytical with this Yukawa form.

The application of the convolution to the Fermi-Breit potential~(\ref{sqqpot}) can be achieved by replacing the gluon propagator $U(r)$ by a ``dressed" one, that is the convoluted propagator~\cite{Brau04} 
\begin{equation}
\tilde U(r)=\frac{1}{r}-\left(\frac{1}{r}+\frac{1}{2\gamma}\right){\rm e}^{-r/\gamma}.	
\end{equation}
This completely removes all singularities in the short-range interactions. For consistency, we apply the same transformation to the confining terms, through the substitution~\cite{Brau04} 
\begin{equation}
ar\rightarrow a\tilde r=ar+\frac{4a\gamma^2}{r}\left(1-{\rm e}^{-r/\gamma}\right)-a\gamma {\rm e}^{-r/\gamma}.
\end{equation}

\subsection{The strong coupling constant}

Lattice QCD clearly shows that the static potential between a quark and an antiquark is compatible with the Cornell form $ar-(4/3)\alpha_Sr$, for $a\approx0.20$ GeV$^2$ and $\alpha_S\approx0.22$ \cite[p. 42]{balirep}. Consequently, the total energy is separated into a confining part, and a ``residual" short-range part -- let us note that this separation is \textit{de facto} performed in our model. The Coulomb term is clearly equal to the lowest order approximation of the Fermi-Breit potential, with a rather small value for $\alpha_S$. Thus, the effective strong coupling constant does not blow up, even in the bound states sector we are dealing with, provided that nonperturbative effects are correctly taken into account. Several attempts have been made in order to include the nonperturbative contributions into the well-known formula giving $\alpha_S(\bm q^2)$, $\bm q$ being the transfered $4$-momentum~\cite{free}. All these approaches qualitatively lead to: 
\begin{equation}\label{asr2}
	\alpha_S(\bm q^2)=\frac{12\pi}{(33-2N_f)\ln\left(\frac{\bm q^2+\xi^2(\bm q^2)}{\Lambda^2}\right)},
\end{equation}
where $\xi(\bm q^2)$ is a monotonic function such that $\xi^2(0)>0$ and $\xi^2(\bm q^2\rightarrow\infty)\rightarrow0$. $N_f$ is the number of quark flavors whose masses are lower than $\bm q^2$, and $\Lambda$ is the famous lambda QCD parameter. However, the explicit formula giving $\xi(\bm q^2)$ is different following the different works. Equation~(\ref{asr2}) states the strong coupling constant remains finite for $\bm q^2=\Lambda^2$ and tends to a maximal value $\alpha_S(0)<1$, in agreement with lattice QCD~\cite{free}. 

A simple, phenomenological, way to mimic the behavior of Eq.~(\ref{asr2}) in position space is to replace $\alpha_S$ by~\cite{sema94}
\begin{equation}\label{asr}
	\alpha_S(r)=\alpha_0\left(1-{\rm e}^{-r/r_c}\right).
\end{equation}
It is nowadays well-established from experimental measurements that $\alpha_S(\bm q^2=m^2_Z)=0.1176\pm0.0020$ \cite{PDG}, with $m_Z\approx 91.19$~GeV. The parameter $r_c$ can consequently be fixed by demanding that $\alpha_S(\bm q^2=m^2_Z)\approx\alpha_S(r=1/m_Z)$. This condition leads to 
\begin{equation}\label{rcdef}
r_c=-\left\{m_Z \ln \left[1-\frac{\alpha_S(\bm q^2=m^2_Z)}{\alpha_0}\right]\right\}^{-1}.	
\end{equation}
The only remaining parameter is $\alpha_0$, that we will fix by a comparison of our results to lattice QCD in the following section. 

\subsection{Summary of the model}

We have now introduced all the ingredients of our model, and one may wonder about the possible double counting of the interactions. Let us sum up the different contributions which are taken into account. 

First of all, the long-range (confining) interactions are encoded in the flux tube. At the lowest order, it reduces to a linear potential, and is thus independent of the features of the confined quarks. As the flux tube is a dynamical object, it carries angular momentum; its dynamical contribution to the total mass, $\Delta M_{rft}$, can be added perturbatively, so defining the perturbative flux tube model. Moreover, the flux tube is assumed to emerge from a flux of exchanged gluons, and retardation effects due to the finite value of light speed are thus present through the term $\Delta M_{ret}$. Then, the spin of the quarks comes into play. It has been shown in Ref.~\cite{Sim1} that the interactions between the spin of a quark and the flux tube can be split in two parts. The first one involves the dynamics of the flux tube through its angular momentum, and leads to the spin-orbit term $\Delta M_{so}$. This term is a relativistic correction that can be thought as a Thomas precession in the color field. The second part does not depend on the flux tube momentum and, after tedious calculations, give rise to the quark self-energy contribution $\Delta M_{qse}$~\cite{Sim1}. Quark self-energy and spin-orbit are thus two different aspects of the interactions induced between the spin of the confined quarks and the flux tube, and both contributions are distinct from dynamical and retardation terms, which are spin-independent. 

As the flux tube model completely neglects the short-range interactions, these have to be added as a Fermi-Breit potential, $V_{oge}(\vec r)$, coming from one gluon exchange diagram. In the same way, the self-energy diagram, which, among other QCD Feynman diagrams, causes the running of the coupling constant, is not contained in the previously mentioned quark self-energy term, which is only linked to the flux tube. Short-range interactions thus complete the model by adding interactions which are all neglected by the flux tube. Finally, mass formula~(\ref{masss}) will be used to compute all the mass spectra in the following. But, $V(\vec r)$, $\alpha_S$, and $U(r)$ will be replaced by $\tilde V_{oge}(\vec r)$, $\alpha_S(r)$ and $\tilde U(r)$ in this last formula, in order to take into account the effective quark size and the running of the coupling constant.

\section{Heavy mesons}\label{hmes}

\subsection{Fitting the parameters}\label{fitting}
\begin{figure}[ht]
\includegraphics*[width=10cm]{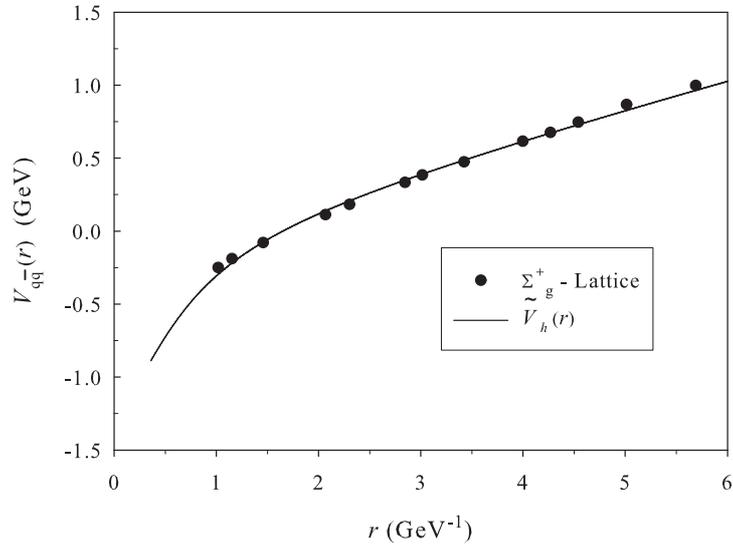}
\caption{Effective quark-antiquark potential computed from lattice QCD data (full circles)~\cite{Juge}. Potential $\tilde V_h(r)$, given by Eq.~(\ref{vtildh}), is plotted with $a=0.185$ GeV$^{2}$, $\alpha_0=0.400$, and $\gamma_h=0.200$ GeV$^{-1}$ (solid line).}
\label{Fig0_2}
\end{figure}

Potential quark models for heavy mesons have proved to be particularly successful for a long time, and still deserve interest because of the new heavy states which are currently being discovered~\cite{old2,russe1,olga,ccbar2}. Moreover, even the complete Fermi-Breit potential, including the relativistic corrections, has recently been validated by lattice QCD calculations~\cite{Koma2}. The various levels of the potential energy between two static quarks have also been computed in lattice QCD~\cite{Juge}. As we already pointed out, the ground state level, denoted as the $\Sigma^+_g$ one in Ref.~\cite{Juge}, appears to be compatible with a standard Cornell potential. But, in order to consistently fit our parameters, we actually have to fit this ground state with the following convoluted potential
\begin{equation}\label{vtildh}
	\tilde V_h(r)=a\tilde r-\frac{4}{3}\alpha_S(r)\tilde U(r).
\end{equation}
It is actually the potential coming from our complete model, where the relativistic corrections vanish in the static quark limit ($m\rightarrow\infty$) since they are expressed in powers of $1/\mu\approx1/m$. As we can see in Fig.~\ref{Fig0_2}, the best fit is obtained with the standard values $a=0.185$ GeV$^2$, $\alpha_0=0.400$, and $\gamma_h=0.200$ GeV$^{-1}$ (the $h$ index denotes the heavy quarks). Such a value for $\alpha_0$ is rather usual in potential models, and defines the value $r_c=0.031$ GeV$^{-1}$ through the relation~(\ref{rcdef}). The parameters of potential~(\ref{vtildh}) are thus completely fixed by lattice QCD. Let us note that $a$ is usually assumed to be around $0.19$~GeV$^2$~\cite[p. 9]{balirep}, in agreement with our fit.

For heavy quarks beyond the static limit, relativistic corrections come into play. They do not demand any new parameter since they are expressed in powers of $1/\mu$, given by Eq.~(\ref{mupftm}). However, two parameters must be added because of the quark self-energy. Following the latest lattice QCD results, we set $f=3.000$ and $\delta=1.000$ GeV~\cite{DiGia1}, although this last parameter has a very little influence (less than $1\%$)~\cite{buis05a}. Finally, the masses of the $c$ and $b$ quarks will be fitted in order to reproduce the $J/\psi$ and $\Upsilon(1S)$ mesons respectively. The values of the different parameters which are used are summarized in Table~\ref{tabparams}.

Before computing detailed mass spectra, it is interesting to make some comments about the influence of the different terms appearing in the mass formula~(\ref{masss}). The numerical values of the different contributions in this last relation are given in Table~\ref{massdec} for several $q\bar q$ states, using the parameters of Table~\ref{tabparams}. As it can be observed, the retardation term, $\Delta M_{ret}$, always brings a negative contribution around $0.1-0.2$ GeV, which is maximal for light mesons and for small quantum numbers. The parameters which have to be used in the case of light mesons will be discussed in Sec.~\ref{lmes}, but performing some calculations are already interesting at this stage in order to understand the evolution of the relativistic corrections when one deals with light instead of heavy quarks. The dynamical contribution of the relativistic flux tube, $\Delta M_{rft}$, also decreases the mass of the mesons, and is logically more and more relevant at high $\ell$, since the flux tube carries orbital angular momentum. A last term decreases the meson mass in every case: The quark self-energy one, denoted as $\Delta M_{qse}$. Its contribution is very important for the $n\bar n$ states (up to $0.5$ GeV), but quickly decreases for heavier mesons. We recall that the symbol $n$ is used to denote both light $u$ or $d$ quarks. It becomes negligible for $b\bar b$ states. The spin-orbit term coming from the flux tube, $\Delta M_{so}$, does not contribute significantly for heavy mesons, but can become relevant for $n\bar n$ states. Finally, the relativistic, short-range, corrections $\left({\rm given\ by} \left\langle V_{oge}(\vec r)+\frac{4}{3} \alpha_S(r) \tilde U(r)\right\rangle \right)$ bring state-dependent contributions, ensuring the mass splitting of the various $J^{PC}$ states. 
 
\subsection{Mass spectra}
\begin{figure}[ht]
\includegraphics*[width=10cm]{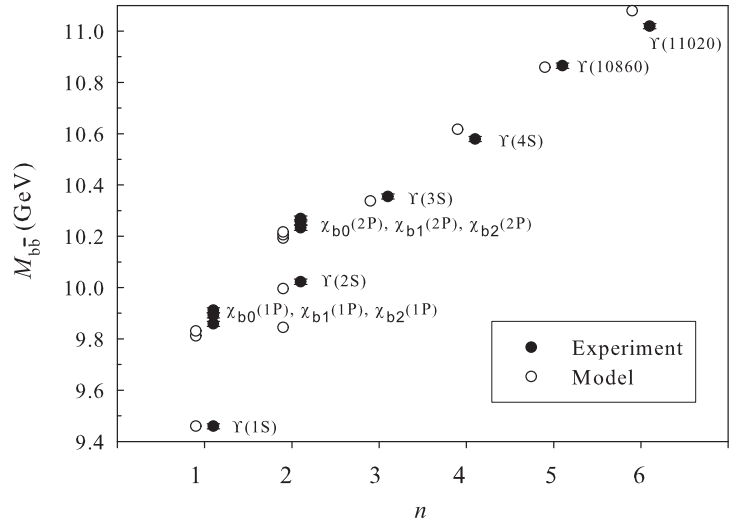}
\caption{$b\bar b$ meson spectrum computed with our model and the parameters of Table~\ref{tabparams} (empty circles). Our results are compared to the experimental data taken from the Particle Data Group~\cite{PDG}.}
\label{Fig1}
\end{figure}

We begin our study by computing the $b\bar b$ meson spectrum, corresponding to the $\Upsilon$ meson family. As it can be observed in Fig.~\ref{Fig1}, the agreement between the experimental data and our spectrum is rather satisfactory, for a $b$ mass given by $m_b=4.785$ GeV. The radial trajectory of the $\Upsilon$ mesons is well reproduced by computing the $1^{--}$, $(n+1)^3S_1$, states . Moreover, our results are compatible with the interpretation of the $\chi_{bJ}$ mesons as the $1^3P_J$ and $2^3P_J$ triplets, although we find them slightly lower than the experimental data.

\begin{figure}[ht]
\includegraphics*[width=10cm]{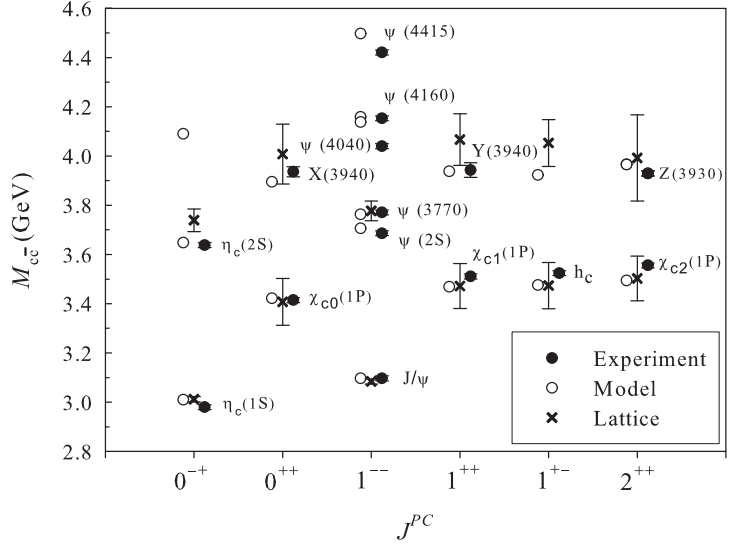}
\caption{Same as Fig.~\ref{Fig1}, but for the $c \bar c$ mesons. More details about the $X$, $Y$, $Z$ states can be found in Ref.~\cite{ccbar2} for example. Lattice results (crosses) are taken from Ref.~\cite{ccbar3}.}
\label{Fig2}
\end{figure}

Among the heavy mesons, charmonia have recently become an important source of discoveries of new experimental states such as the $X(3940)$, $Y(3940)$, \textit{etc.} (see Refs.~\cite{ccbar1,newhadr,newhadr2}). That is why a computation of the $c\bar c$ spectrum within the framework of our generalized flux tube model is of particular interest. The results are plotted in Fig.~\ref{Fig2}. It appears that our spectrum fits the data with a reasonable agreement, excepted for the $\psi(4040)$ and $\psi(4415)$ mesons, which are expected to be a $3^3S_1$ and a $4^3S_1$ state respectively. They are overestimated in our model, as it is the case in a previous study relying on a Cornell potential, thus sharing some similarities with this work~\cite{char}. The $\psi(3770)$ is however well matched by a $1^3D_1$ state. We can also observe that the predictions of our model concerning the $2^3P_J$ $c\bar c$ multiplet, for which no firm experimental candidate is known, are compatible with the lattice QCD study of Ref.~\cite{ccbar3}, as well as with other studies on the charmonium spectrum in lattice QCD (see Ref.~\cite{latticcbar} for a review on the subject). 


It is now interesting to focus on the recently discovered $X(3940)$, $Y(3940)$, and $Z(3930)$ states. It has been argued in the literature that the $Y(3940)$ and the $Z(3930)$ are the $\chi_{c1}(2P)$~\cite{ccbar2} and the $\chi_{c2}(2P)$~\cite{newhadr2} respectively. Our results agree with that suggestion. But, it has also been suggested that the $X(3940)$ is the $\eta_c(3S)$~\cite{ccbar2,newhadr}. However, as it has already been pointed out, quark models predict a too high mass for that state~\cite{newhadr2}. The most satisfactory assignment for the $X(3940)$ from the point of view of our flux tube model is that this state is the $\chi_{c0}(2P)$, in agreement with our spectrum (see Fig.~\ref{Fig2}). Consequently, we suggest that these three $X$, $Y$, $Z$ states could be identified with the $\chi_{cJ}(2P)$ triplet.      

\section{Light mesons}\label{lmes}

The model we presented in Sec.~\ref{model} has the nice feature that it is well-defined for light particles, even for massless ones. That is why the computation of light meson spectra within our framework seems to be relevant.

\subsection{Fitting the parameters}
It is worth discussing a bit about how the parameters of the model should be modified in the light meson sector. Rigorously, an effective model of light mesons should take into account the possible decays and coupling with other channels for any given state. This is clearly not the case in our model, as in most of the quark models of mesons. Generally, this problem is bypassed by fitting the parameters on the experimental spectrum in order to have an agreement as satisfactory as possible between theory and experiment. Up to now, we did not fit the parameters on the experimental data, but rather on lattice QCD. Actually, our way of proceeding is indirectly based on experiment through lattice QCD. It is indeed known that this last approach correctly reproduces the heavy meson spectra (see for example Ref.~\cite{latticcbar}). But, its advantage is that it can also provide us with an estimation of the quark-antiquark potential energy and of the values of $f$ and $\delta$. It is thus more straightforward to fit our potential on the corresponding lattice QCD curve than on many experimental data, knowing that lattice QCD correctly reproduces the experimental spectra. By this way, we are led to particular values for $a$, $\alpha_0$, and $\gamma_h$, but the quark masses $m_b$ and $m_c$ have still to be adjusted on the experimental masses, as it is usually done in potential models. 

Neither the flux tube energy density $a$ nor the and the coupling constant at zero momentum, $\alpha_0$, are expected to vary with the quark mass, thus the same values than in the case of heavy mesons can be kept to describe light mesons. The same remark holds for $f$ and $\delta$. 

We turn now our attention to the light quark masses. The mass of the $n$ quark, is fixed at $m_n=0$ for future convenience, in very good approximation of the current experimental data, stating that $1.5$ MeV$<m_u,m_d<7.0$ MeV~\cite{PDG}. Let us note again that no distinction has to be made between $u$ and $d$ quarks because our model is isospin-independent. The mass of the $s$ quark is fitted to reproduce the $\phi(1020)$, which is largely accepted as a pure $s\bar s$ state (see for example Ref.~\cite{old2}). We obtain $m_s=0.202$~GeV.
    
Finally, we have to ask whether the effective size $\gamma_h$ should be modified or not in the light quark sector. Intuitively, one can imagine that a heavy quark is closer of the classical picture of a pointlike particle than a light one. That is why it seems justified to assume that $\gamma_l>\gamma_h$, where the $l$ index stands for light quarks. After a fit on the experimental data, we find indeed that the best agreement with our model is achieved for $\gamma_l=0.940$ GeV$^{-1}$. That is thus the value that we will choose. Interestingly, such a value corresponds to the effective quark size which is generated by instanton-induced interactions, as we will mention in the next section. 

\subsection{Instanton-induced forces}

The lightest mesons are the $n\bar n$ ones. Since we set $m_n=0$, it is worth mentioning that the results are exactly independent of the value of $\delta$. An additional physical mechanism has to be taken into account in the $n\bar n$ mesons, which is the instanton-induced forces. Instantons are classical solutions of the euclidean equations of motion of QCD, which provide informations on the nontrivial vacuum structure in QCD. In a light meson, it has been shown that instantons induce forces between the quark and the antiquark (see Ref.~\cite{instanton3} for a review). Such forces can be included as a potential in quark models, and only give a nonzero contribution for the $^1S_0$ $n\bar n$ states~\cite{instanton}. A proper calculation of the influence of these effects on the light meson mass spectra is rather complicated, and was performed in Ref.~\cite{instanton} with a set of parameter nearly identical to ours, denoted as Model~III. The main change resides in the value of the string tension $a$. But, if $\Delta M_{ins}$ is the instanton contribution, $\Delta M_{ins}/\sqrt{a}$ is universal since we deal with massless quarks ($m_n=0$). We can then compute from Ref.~\cite{instanton} that 
\begin{eqnarray}\label{inst}
	\Delta M^\pi_{ins}&=&-0.438\ {\rm GeV},\ \Delta M^{\pi(1300)}_{ins}=-0.205\ {\rm GeV},\ \Delta M^{\pi(1800)}_{ins}=-0.119\ {\rm GeV},\nonumber\\
	\Delta M^\eta_{ins}&=&-0.042\ {\rm GeV}.
\end{eqnarray}
Instanton-induced forces do not act on other states than the $\pi$ and $\eta$ ones. They are thus very different in nature than the quark self-energy contribution, which acts on every $q\bar q$ state, although both terms are related to quark interactions with a nontrivial configuration of the gluonic field. Actually, the different status of these interactions could be understood by considering the different length scales appearing in the model. Indeed, the flux tube configuration emerges at large distances and give rise to numerous interactions, which were presented in Sec.~\ref{model}, including quark self-energy. Instantons are however particular configurations of the gluonic field whose typical size is well smaller than the typical size of a hadron (roughly related to the size of the flux tube)~\cite{instanton2}, and they thus lead to different phenomenological implications. 

It can be shown that instantons give an effective size to light quarks, which is generally computed to be around $1$ GeV$^{-1}$~\cite{DiGia1,instanton3,instanton2}. This effective size is actually linked to the nonzero value of the mixed quark-gluon condensate $g\left\langle\bar q \, \sigma_{\mu\nu}F^{\mu\nu}q \right\rangle$ (see for example Refs.~\cite{DiGia1,instanton2}). Remarkably, this effective size is close to the fitted value that we use, namely $\gamma_l=0.940$ GeV$^{-1}$. The interpretation of our phenomenological effective size in terms of the effective size generated by instantons in the light quark sector seems thus rather natural. It is worth mentioning that, rigorously, instantons also confer a constituent mass to the light quarks, clearly because of the breaking of chiral symmetry. But, this constituent mass (of order $10$ MeV) is small enough with respect to the dynamical mass $\mu\approx300$ MeV generated by the confinement~\cite{instanton}, and then can be neglected.   

\subsection{Mass spectrum} 

\begin{figure}[ht]
\includegraphics*[width=10cm]{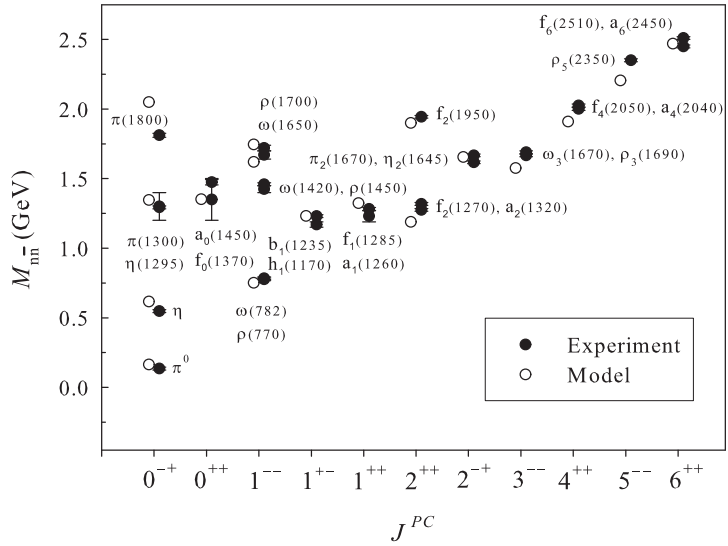}
\caption{Same as Fig.~\ref{Fig1}, but for the $n \bar n$ mesons. The instanton contribution~(\ref{inst}) has been added to the $\pi$ and $\eta$ states.}
\label{Fig3}
\end{figure}

The $n\bar n$ spectrum is plotted in Fig.~\ref{Fig3} and compared with the experimental data. As our model neglects the isospin, we only get one point corresponding to each isospin doublet. We see that our spectrum is in correct agreement with these data, excepted for the $\pi(1800)$ -- but the experimental evidence for this last state is weak. It is interesting to notice that the mass we find for the lowest $0^{++}$ state is $1.351$ GeV, close to the $f_0(1370)$. Our model thus support the idea that the $f_0(1370)$ is a dominantly $n\bar n$ state, as it is argued in Ref.~\cite{mix1}. Among the numerous $2^{++}$ mesons, the $f_2(1950)$ seems to be a pure $n\bar n$ state, with quantum numbers $2^3P_2$.     

It is worth mentioning that no scalar state is present between $0.616$ GeV and $1.346$ GeV -- corresponding to the $\eta$ meson and to the $\eta(1295)$ and $\pi(1300)$ respectively -- in our $n\bar n$ spectrum. However, recent experimental results have provided evidence for low-lying states with quantum numbers $0^{++}$ and masses in this range, namely the $f_0(600)$, $\kappa(800)$, $f_0(980)$, and $a_0(980)$ (useful references can be found in~\cite{tetra2}). The recent detailed study of Ref.~\cite{tetra2} suggests that these states are not $q\bar q$ mesons, but rather that they are part of the lowest-lying scalar tetraquark nonet, in agreement with earlier works on this subject~\cite{tetra}. Following these results, we thus conclude that these light experimental scalar states cannot be reproduced by our model, since we only deal with $q\bar q$ mesons, and not with $qq\bar q\bar q$ tetraquarks. 

A correct description of $s\bar s$ mesons cannot be achieved without including a mixing with $n\bar n$ mesons~\cite{old2}. As we neglected such couplings, a detailed $s\bar s$ spectrum will not be computed. We obtain $m_s=0.202$ GeV. Then, the lowest $0^{++}$ state ($1^3P_0$) has a mass given  by $1.528$ GeV, which is near the $f_0(1500)$. This state could then be a mostly $s\bar s$ state, in agreement with Ref.~\cite{mix1}. Moreover, the $2^3P_2$ and $1^3F_2$ states have the following mass respectively: $2.062$ and $2.226$ GeV, which suggest that the $f_2(2010)$ and $f_2(2300)$ are principally $s\bar s$ mesons.

\begin{figure}[ht]
\includegraphics*[width=10cm]{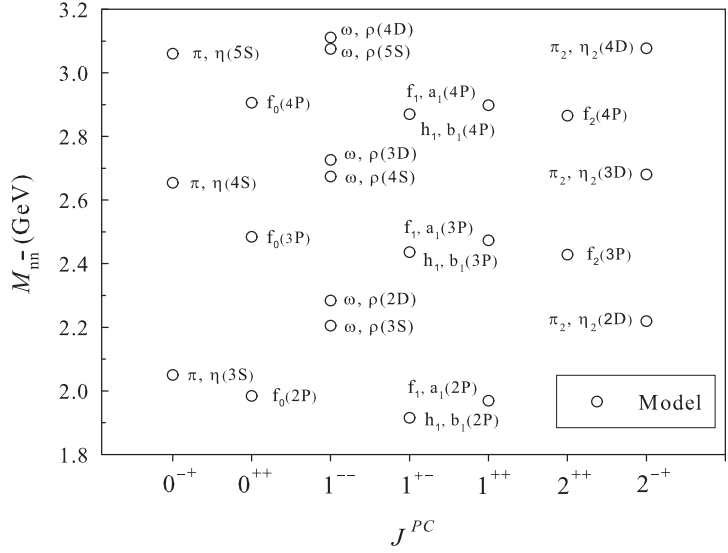}
\caption{Theoretical predictions for the masses of some highly excited $n\bar n$ mesons (empty circles). Parameters of Table~\ref{tabparams} were used.}
\label{Fig3_2}
\end{figure}

Let us end this analysis of light mesons by a computation of some highly excited $n\bar n$ states. The interest of performing such a calculation actually comes from future experiments like PANDA and GLUEX, which will scan the hadron spectrum in the range $1.5-3.0$~GeV. One can hope that these experiments will not only find new mesons, but also detect more intriguing particles like glueballs or hybrid mesons. If we assume that a hybrid meson is a $q\bar q g$ bound state, it can be shown that the lowest-lying ones should have the following quantum numbers~\cite{kalaqqg}: $J^{PC}=0^{\pm+},\, 1^{\pm+},\, 1^{\pm-},\, 2^{\pm+}$. Among these quantum numbers, only $1^{-+}$ is an exotic state, that is a state with quantum numbers that a usual meson cannot possess. The discovery of such an exotic state could be the unambiguous signature of a hybrid meson. On the contrary, the other possible $J^{PC}$ are accessible for usual mesons. It seems thus interesting for the interpretation of the future experimental results to compute with our model the masses of the $0^{\pm+},\, 1^{++},\, 1^{\pm-},\, 2^{\pm+}$ $n\bar n$ states in the interval $2-3$ GeV, in order to suggest energy levels around which an usual meson should be located. This could eventually lead to identify hybrid mesons with non exotic quantum numbers at PANDA or GLUEX. Results are plotted in Fig.~\ref{Fig3_2}. Since our model neglects the isospin interactions, this plot has to be seen as predictions giving masses around which an experimental $n\bar n$ state should be located rather than accurate predictions of meson masses. As an example of a possible application of Fig.~\ref{Fig3_2}, we can look at the $1^{--}$ states. Following our predictions, no $n\bar n$ states with $J^{PC}=1^{--}$ should be present in the interval $2.4-2.6$ GeV. This defines a particular mass range in which it should be interesting to search for a hybrid meson in future experiments. Indeed, if a $1^{--}$ state with isospin $1$ was found in this mass range for example, it could be a serious candidate for a $n\bar n g$ hybrid meson: An isospin equal to $1$ forbids a possible $s\bar s$ state, and no glueball with such $J^{PC}$ is expected at such a low mass, as we will see in the following.        

\section{Glueballs}\label{gb}

\subsection{Parameters}
The model we present in this study, although being rather simple, can describe with a satisfactory agreement the meson spectra, even in the light sector. The next step is to wonder whether such a flux tube approach can be applied to glueballs or not. The subject is still a matter of controversy. Let us review the different parts of our model, assuming in a first time that a glueball can be seen as a bound state of two constituent massless gluons ($m_g=0$). The situation is then very similar to light quarks. The dynamical flux tube and retardation terms are only defined by the flux tube itself and do not depend on the nature of the confined particles. Moreover, we have recently shown that the spin-orbit term arising from the flux tube has the same form for particles of arbitrary spin~\cite{buis07_so}. On the contrary, the one-gluon-exchange potential should be modified, since the effective potential emerging from the Feynman diagrams involving a pair of gluons at tree level is different of the one between a quark and an antiquark. It has been computed in Ref.~\cite{oge}, and reads
\begin{eqnarray}\label{sggpot}
    V_{oge}(\vec r\, )&=&-3\alpha_S\left\{U-\frac{1}{\mu^2}\left(\frac{1}{2}-\frac{13}{48}\vec S^{\, 2}\right)\Delta U+\frac{3}{2\mu^2}\vec L\cdot\vec S \ \frac{U'}{r}\right.\nonumber\\
    &&\left.+\frac{1}{6\mu^2}\left(U''-\frac{U'}{r}\right)\left[\vec S^2-3\frac{(\vec S\cdot\vec r)^2}{r^2}\right]\right\},
\end{eqnarray} 
where $U(r)$ is again the gluon propagator in position space, and where $\mu$ is the dynamical gluon mass, which typically has a value around $0.5-0.6$ GeV, in agreement with other effective approaches~\cite{new,silva}. A last modification occurs in the self-energy term: Strong theoretical and phenomenological arguments indicate indeed that gluons do not bring any contribution of self-energy \cite{kaid}. This term consequently vanishes in glueballs. For what concerns the instanton contribution, we can mention a previous study stating that instanton-induced interactions could significantly contribute in the $0^{\pm+}$ glueballs~\cite{inst}. But, to our knowledge, the inclusion of such interactions has not been elucidated yet within the framework of constituent gluon models and consequently, we will not consider any instanton-like contribution in glueballs. 

\begin{figure}[ht]
\includegraphics*[width=10cm]{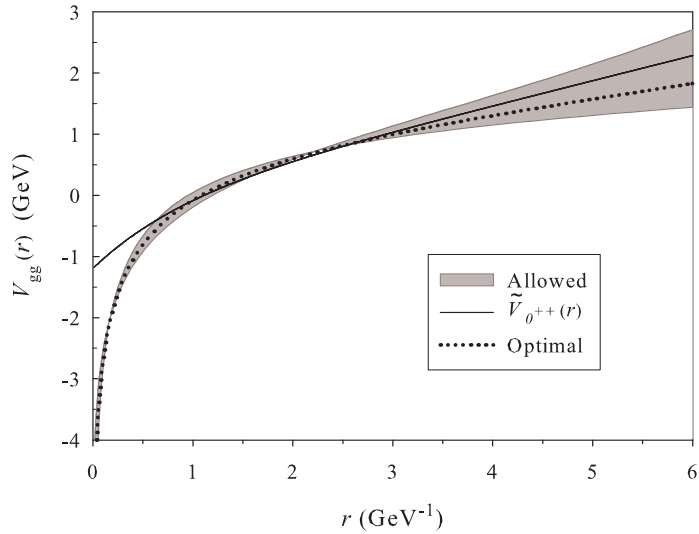}
\caption{Effective gluon-gluon potential computed from lattice QCD data~\cite{gluwf} concerning the $0^{++}$ glueball~\cite{glue2}. The optimal potential is plotted with a dotted line, but the error bars on the lattice results allow the true effective potential to be located in the gray area. Potential $\tilde V_{0^{++}}(r)$, given by Eq.~(\ref{vtild}), is plotted with $a_g=(9/4)\, 0.185$ GeV$^{2}$, $\alpha_0=0.400$, and $\gamma_g=0.940$ GeV$^{-1}$ (solid line).}
\label{Fig0_1}
\end{figure}

Finally, the basic remaining question to answer to is: Can glueballs (at least the low-lying ones) be described by a bound state of two constituent gluons? Roughly speaking, we gave an affirmative answer in Ref.~\cite{glue2}. We showed in this last work that the lattice QCD mass and wave function of the lowest lying $0^{++}$ glueball~\cite{gluwf} are compatible with those of a bound state of two constituent massless particles, interacting via a Cornell potential. The results of Ref.~\cite{glue2} are plotted in Fig.~\ref{Fig0_1}. The optimal potential, obtained by using the optimal values of the lattice QCD data, clearly exhibits a confining long-range part, and a rapidly decreasing short-range part. The errors on these lattice data allow the ``true" potential to be located between two extremal curves, in the gray area. The curve we get in Ref.~\cite{glue2} can be compared to the potential for the $0^{++}$ glueball in the present flux tube model. This state is a $\ell=S=0$ one. Consequently, only the contact interaction and the retardation term have to be added to the convoluted linear-plus-Coulomb potential. Using the approximation~(\ref{retapp}) for massless particles, we find that the potential is of the form
\begin{equation}\label{vtild}
	\tilde V_{0^{++}}(r)=a_g\, \tilde r-3\, \alpha_S(r)\left[\tilde U(r)+\frac{{\rm e}^{-r/\gamma_g}}{4\mu^2\gamma_g^3}\right]-\frac{3a_g}{8\mu}.
\end{equation}
In this potential, the tension $a_g$ for the flux tube in a glueball is related to the corresponding one in a meson, simply denoted as $a$, through a scaling law $a_g={\cal C}a$. The constant ${\cal C}$ is generally assumed to be given by $9/4$ (Casimir scaling) \cite{scaling}, or by $3/2$ (square root of Casimir scaling) \cite{bagm}. We assume here the Casimir scaling hypothesis, which has been confirmed by lattice QCD calculations as well as by effective models~\cite{new,scaling,scabicu}. The choice ${\cal C}=3/2$ actually emerges from bag models-inspired approaches, that are not considered in this work. It is worth noting that a $3$ factor is now present before $\alpha_S(r)$, which corresponds to the value of the color operator in the one-gluon-exchange diagram between two gluons. Moreover, we showed in Ref.~\cite{hyb1} that, in the ground state,
\begin{equation}\label{vtild2}
	\mu\approx\sqrt{\frac{3\pi a_g}{16}}.
\end{equation}

The potential defined by relations~(\ref{vtild}) and (\ref{vtild2}) can be compared to the results coming from lattice QCD: For $\alpha_0=0.400$ and $\gamma_g=\gamma_l=0.940$ GeV$^{-1}$, it is maximally located in the gray area, as it can be observed in Fig.~\ref{Fig0_1}. Interestingly, the effective gluon size is compatible with the one of massless quarks. Assuming the Casimir scaling hypothesis, the same parameters than for light mesons can thus be used. 

\subsection{Mass spectrum}
\begin{figure}[ht]
\includegraphics*[width=10cm]{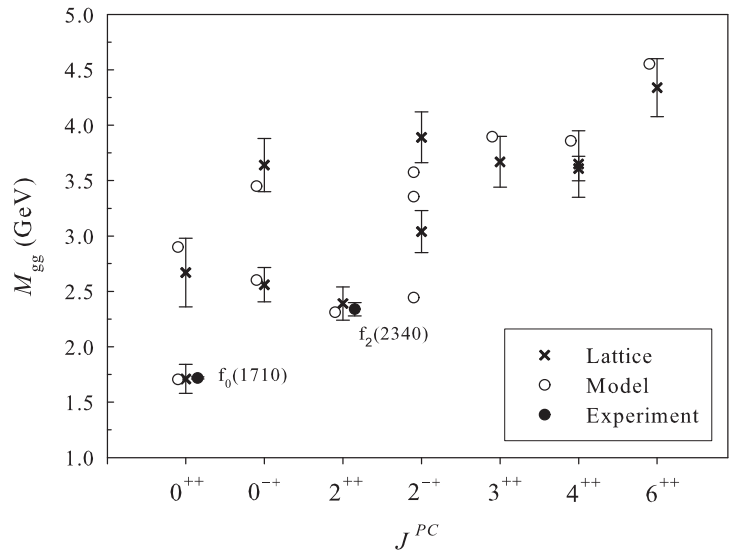}
\caption{Graphical representation of the results presented in Table~\ref{resglu}. The spectrum obtained with our model (empty circles) is compared to lattice QCD computations (crosses) and to possible experimental candidates mentioned by the Particle Data Group (full circles)~\cite{PDG}.}
\label{Fig4}
\end{figure}

The lowest-lying glueball states are given in Table~\ref{resglu}, and plotted in Fig.~\ref{Fig4}. Our results are always located in the error bars of lattice QCD~\cite{latti2,latti1,latti3,latti4}, excepted for the $2^{-+}$ states. However, more states are present in our spectrum than in lattice QCD, i.e. $J=1$ states. Lattice calculations seem indeed to rule out the presence of $1^{-+}$ and $1^{++}$ states below 4 GeV~\cite{latti2}. This can be qualitatively understood in terms of interpolating operators of minimal dimension, which can create glueball states, with the expectation that higher dimensional operators create higher mass states: The lowest states $0^{++}$, $2^{++}$, $0^{-+}$ and $2^{-+}$ are produced by dimension$-4$ operators, while $1^{++}$ and $1^{-+}$ states are respectively produced by dimension$-5$ and dimension$-6$ operators~\cite{latti1,latti2}. Nevertheless, we can observe in Table~\ref{resglu} that our model predicts the existence of $1^{-+}$ and $1^{++}$ states around 3 GeV. To our knowledge, no such experimental candidate has been found yet. The presence of these states in our model may actually be due to the use of spin degrees of freedom instead of helicity ones, which could be the correct ones if the constituent gluons are massless. Let us note that this point is not always accepted in the literature: Many approaches involve indeed massive valence gluons~\cite{Brau04,corn,hou}. An intuitive argument is the following. If the spin is taken as internal degree of freedom, the Pauli principle states that $S$ has to be even ($0$ or $2$) for even $\ell$, and odd (only $1$) for odd $\ell$. But, the combination of two gluons with the helicity $h$ as degree of freedom gives the states $h=0^A$, $0^S$ and $2$, where $0^{S(A)}$ is a (anti)symmetric configuration. For even $\ell$, the possible helicities are $0^S$ or $2$, as for the spin. But, for odd $\ell$, $h=0^A$ is the only possibility. The differences between spin and helicity degrees of freedom will thus mainly manifest for odd $\ell$. Interestingly, the $2^{-+}$ states, which are missed in our model, are $\ell=1$ states. We thus think that, although the rough picture as a two-gluon bound state interacting via a Cornell potential seems to be valid, a more careful study of the relevant degrees of freedom for the constituent gluons is the key to a possible fully quantitative application of potential models to glueballs. The presence of spurious states around 3 GeV in our model could indicate what are the limits of the present approach.

From the previous discussion, we can guess that the $\ell=0$ glueball states are correctly described within our model, whether spin or helicity have to be used. These states are the $0^{++}$ and $2^{++}$ ones, which are compatible with the $f_0(1710)$ and the $f_2(2340)$ respectively. Consequently, our results support the identification of the $f_0(1710)$ with a pure glueball state, in agreement with Ref.~\cite{mix1}. The $f_2(2340)$ should also be seen as a serious candidate for the tensor glueball.

\section{Scalar glueball and meson mixing}\label{mixi}
In order to understand the numerous light $0^{++}$ states which have been experimentally detected, it was firstly suggested in Ref.~\cite{Amsler} that these scalar states could be interpreted as mixings between pure $\left|gg\right\rangle$, $\left|n\bar n\right\rangle$ and $\left|s\bar s\right\rangle$ states. Many works have since been devoted to the study of the mixing between scalar mesons and glueball (see Refs.~\cite{mix1,Close96,Close02,Closeaf,mix2,berna}), with different results following the approaches. In our framework, the masses of the pure $0^{++}$ states are 
\begin{equation}
	M_{gg}=1.705\, {\rm GeV},\ M_{s\bar s}=1.528\, {\rm GeV},\ M_{n\bar n}=1.351\, {\rm GeV}.
\end{equation}
The structure of mixed states can be investigated thanks to a mixing matrix~\cite{Close96}, which, following the lattice study of Ref.~\cite{mix2}, is of the form
\begin{equation}\label{mixing}
	\begin{pmatrix}
	M_{gg} & \lambda & \sqrt{2}\, r\, \lambda\\
	\lambda & M_{s\bar s} & 0\\
	\sqrt{2}\, r\, \lambda & 0 & M_{n\bar n}
	\end{pmatrix}, 
\end{equation}
with $\lambda=0.064\pm0.013$ GeV and $r=1.198\pm0.072$~\cite{mix2}. In this last study, the mixing between pure states was used to understand the structure of the $f_0(1370)$, $f_0(1500)$, and $f_0(1710)$ mesons. However, following our previous conclusions, we assume that the $f_0(1710)$ is the pure glueball state, and that the $f_0(1370)$ is the pure $n\bar n$ state. 

We would like to add in the game the recently observed $f_0(1810)$~\cite{abli}, which can be considered also as a good glueball candidate~\cite{bicu}. It is worth mentioning that this interpretation is not unanimous: In some studies, the $f_0(1810)$ is interpreted as a mainly $s\bar s$ hybrid meson state~\cite{chine}. However, hybrid mesons are not included in the present discussion. As we only consider the mixing between usual mesons and glueball, it is thus interesting to determine whether the $f_0(1810)$, whose mass is $1.812\pm0.044$ GeV~\cite{abli}, and the $f_0(1500)$, whose mass is $1.507\pm0.005$ GeV~\cite{PDG}, can be reproduced by the simple mixing matrix~(\ref{mixing}). To do that, we begin by fixing the ratio $r$ to the value $1.198$, as it was computed by lattice QCD calculations. Since we are interested in different states that in Ref.~\cite{mix2}, we will allow the parameter $\lambda$ to be slightly different that the value $0.064$ GeV. Interestingly, for $\lambda=0.104$ GeV ($35\%$ higher than the maximal value of Ref~\cite{mix2}), we find three eigenstates defined as follows:
\begin{eqnarray}
M_1&=&1.811\, {\rm GeV},\ \left|\psi_1\right\rangle	=-0.883 \left|gg\right\rangle-0.325\left|s\bar s\right\rangle-0.338 \left|n\bar n\right\rangle,\\
M_2&=&1.502\, {\rm GeV},\ \left|\psi_2\right\rangle	=+0.237 \left|gg\right\rangle-0.931\left|s\bar s\right\rangle+0.277 \left|n\bar n\right\rangle,\\
M_3&=&1.272\, {\rm GeV},\ \left|\psi_3\right\rangle	=-0.405 \left|gg\right\rangle+0.164\left|s\bar s\right\rangle+0.899 \left|n\bar n\right\rangle.
\end{eqnarray}

We see that, for this particular value of $\lambda$, $M_1$ is compatible with the mass of the $f_0(1810)$, which could thus be a mainly glueball state. Moreover, $M_2$ can be identified with the $f_0(1500)$, with a dominant $s\bar s$ component. An additional light $0^{++}$ meson is now present, which is still compatible with the $f_0(1370)$, because of its large error bar. Actually, the mass of this state have been measured in a rather large range, from $1.2$ to $1.5$ GeV~\cite{PDG}. Only keeping in mind the latest data, the pure $n\bar n$ state we previously found is merely compatible with the state of Ref.~\cite{abli2}, which was detected at a mass of $1.350\pm0.050$ GeV. That clearly corresponds to the $f_0(1370)$. But, in Ref.~\cite{abli1}, a state was detected with a mass equal to $1.265\pm0.065$ GeV. This state was also identified with the $f_0(1370)$, but its mass is rather compatible with our $M_3$ state. Our approach then suggests that two states hide behind the current $f_0(1370)$. The heaviest one is compatible with a mass of $1.370$ GeV, while the lightest one has a mass around $1.270$ GeV. Further experimental studies should confirm this hypothesis. 

It is worth mentioning that, for $r=1.270$ and $\lambda=0.077$ GeV, which are the largest values allowed following the lattice study of Ref.~\cite{mix2}, one finds 
\begin{eqnarray}
M_1&=&1.774\, {\rm GeV},\ \left|\psi_1\right\rangle	=-0.911 \left|gg\right\rangle-0.285\left|s\bar s\right\rangle-0.298 \left|n\bar n\right\rangle,\\
M_2&=&1.509\, {\rm GeV},\ \left|\psi_2\right\rangle	=+0.231 \left|gg\right\rangle-0.952\left|s\bar s\right\rangle+0.202 \left|n\bar n\right\rangle,\\
M_3&=&1.300\, {\rm GeV},\ \left|\psi_3\right\rangle	=-0.341 \left|gg\right\rangle+0.115\left|s\bar s\right\rangle+0.933 \left|n\bar n\right\rangle.
\end{eqnarray}
The previous conclusions remain qualitatively correct without fitting the parameter $\lambda$, even if the agreement with the experimental masses is less accurate.

\section{Conclusions}\label{conc}

We presented in this work a generalized version of the relativistic flux tube model. Not only the dynamical contribution of the flux tube were taken into account, but also the retardation effects and the spin-orbit interactions generated by the flux tube itself. We also added short-range interactions emerging from one-gluon-exchange process, and for the meson case, quark-self-energy and instanton-induced forces. All these physical ingredients allow to drop the inclusion of an arbitrary constant, which is usually used in potential models to fit the absolute mass scale of the mass spectrum. Moreover, all the parameters appearing in our model can be fitted on lattice QCD results, which strongly reduces its arbitrariness. 
   
We firstly computed the heavy meson spectra. The $b\bar b$ states are globally correctly reproduced, as well as the $c\bar c$ states. In particular, our results suggest to identify the three new states $X(3940)$, $Y(3940)$, and $Z(3930)$, with the triplet $\chi_{cJ}(2P)$. Let us note that a previous study already suggested that $Y(3940)$ and $Z(3940)$ are the $\chi_{c1}(2P)$ and $\chi_{c2}(2P)$~\cite{ccbar2}. 

Since the relativistic flux tube model is well defined for light quarks, and because the results in the heavy meson sector were encouraging, we applied it to the light meson case. The $n\bar n$ spectrum is rather well described, excepted that our model misses the isospin splittings, which we do not take into account in this approach. We compute that a $0^{++}$ state is present at a mass of $1.351$ GeV, in agreement with the $f_0(1370)$~\cite{abli2}. Following our results, the $f_2(1950)$ also seems to be a pure $n\bar n$ state. As we did not include the possibility of mixing between $n\bar n$ and $s\bar s$ states, only three states, which can be seen as nearly pure $s\bar s$, can be described: the $\phi(1020)$, the $f_2(2010)$, and the $f_2(2300)$. 


Finally, we tried to extend the model to glueballs. Assuming the Casimir scaling hypothesis, we got a spectrum which is mostly located in the error bars of the lattice QCD one. However, spurious states appear in our spectrum around $3$ GeV, which are not present in lattice QCD. It consequently appears that the application of potential models to glueballs requires a particular care. Indeed, the gluons are massless, and consequently, their internal degree of freedom should be helicity. However, the relativistic corrections we use are computed as if a gluon was a massive spin-$1$ particle, whose mass is the dynamical mass $\mu=\left\langle \vec p^{\, 2}\right\rangle$. We then obtain expressions involving spin degrees of freedom. As we argued in this work, a careful study of the states obtained either with helicity or spin should lead to a better understanding of the domain of validity of potential models. We leave such an analysis for future works.

Our results concerning glueballs can fortunately be assumed to be valid for the lowest lying $0^{++}$ and $2^{++}$ states. Our model then suggests that the $f_0(1710)$ is a pure glueball and that the $f_2(2340)$ is a very good candidate for the tensor glueball. Moreover, if we add the possibility of a mixing between scalar mesons and glueball, we find that our model is coherent with the $f_0(1810)$ and the $f_0(1500)$ as dominantly glueball and $s\bar s$ states respectively. We also predict a third state around $1.270$ GeV, that could be a new scalar state, mostly $n\bar n$, probably already detected~\cite{abli1} but wrongly identified with the $f_0(1370)$.   

\acknowledgments
The author is grateful to Dr Claude Semay for his constant interest in this work and for his advices, and to Dr Nicolas Boulanger for useful discussions. He also thanks the FNRS for financial support.

\begin{table}[ht]
	\centering
		\begin{tabular}{lc|lc}
\hline\hline
$a$ (GeV$^{2}$) & 0.185 & $\ m_b$ (GeV) & 4.785 \\
$\alpha_0$ & 0.400& $\ m_c$ (GeV) & 1.445\\
$\gamma_h$ (GeV$^{-1}$)  & 0.200 & $\ m_s$ (GeV) & 0.202\\
$\gamma_l=\gamma_g$ (GeV$^{-1}$)  & 0.940	& $\ m_n$ & 0.000 \\
$f$ & 3.000 & $\ m_g$ & 0.000 \\
$\delta$ (GeV) & 1.000	& &\\
\hline\hline			
		\end{tabular}
		\caption{Numerical values of the parameters which are involved in our computations. $m_n$ commonly denotes the mass of the $n$ quark, which commonly denotes $u$ or $d$ quarks.}
		\label{tabparams}
\end{table}

\begin{table}[ht]
	\centering
		\begin{tabular}{cccccccc}
		\hline\hline
$q\bar q$ & $(n+1)^{2S+1}L_J$ & $M_0$ & $\Delta M_{rft}$ & $\Delta M_{ret}$  & $\Delta M_{so}$ & $\Delta M_{qse}$ &  $\left\langle V_{oge}(\vec r)+\frac{4}{3} \alpha_S(r) \tilde U(r)\right\rangle $\\
\hline 
$b\bar b$& $1^3S_1$ & 9.572 & 0 & -0.127 & 0 & -0.002 & 0.018 \\
         & $2^3S_1$ & 10.076 &0 & -0.086 & 0 & -0.002 & 0.008 \\
         & $1^3P_0$ & 9.939 & -0.002 & -0.098 & 0.005 & -0.002 & -0.028 \\
         & $1^3D_1$ & 10.180 & -0.003 & -0.084 & 0.005 & -0.002 & -0.016 \\
$c\bar c$& $1^3S_1$ & 3.240 & 0 & -0.166 & 0 & -0.031 & 0.053 \\
         & $2^3S_1$ & 3.814 & 0 & -0.115 & 0 & -0.029 & 0.037 \\
         & $1^3P_0$ & 3.645 & -0.009 & -0.131 & 0.027 & -0.030 & -0.080 \\
         & $1^3D_1$ & 3.937 & -0.017 & -0.112 & 0.028 & -0.030 & -0.042 \\
$n\bar n$& $1^3S_1$ & 1.404 & 0 & -0.195 & 0 & -0.533 & 0.075 \\
         & $2^3S_1$ & 2.067 & 0 & -0.130 & 0 & -0.352 & 0.034 \\
         & $1^3P_0$ & 1.847 & -0.054 & -0.160 & 0.217 & -0.395 & -0.105 \\
         & $1^3D_1$ & 2.208 & -0.088 & -0.135 & 0.178 & -0.327 & -0.076 \\
\hline		\hline
		\end{tabular}
		\caption{Numerical evaluation of the different terms appearing in the mass formula~(\ref{masss}) for some $q\bar q$ states. Parameters of Table~\ref{tabparams} are used. The nonzero values are given in GeV.}
		\label{massdec}
\end{table}

\begin{table}[ht]
	\centering
		\begin{tabular}{ccccc}
		\hline\hline
$(n+1)^{2S+1}L_J$ & $J^{PC}$ & Model & Lattice & Experiment \\
\hline 
$1^1S_0$ & $0^{++}$ & 1.705 & $1.710\pm0.050\pm0.080$ \cite{latti2} & $1.718\pm 0.006$~\cite{PDG}\\
$1^5S_2$ & $2^{++}$ & 2.311	& $2.390\pm0.030\pm0.120$ \cite{latti2} & $\ 2.339\pm 0.060$~\cite{PDG}\\
$1^3P_0$ & $0^{-+}$ & 2.602 & $2.560\pm0.035\pm0.120$ \cite{latti2} & \\
$1^3P_1$ & $1^{-+}$ & 2.594 & & \\
$1^3P_2$ & $2^{-+}$ & 2.443 & & \\ 
$2^1S_0$ & $0^{++}$ & 2.899 & $2.670\pm0.180\pm0.130$ \cite{latti1} & \\
$1^5D_1$ & $1^{++}$ & 3.152 & & \\
$2^3P_2$ & $2^{-+}$ & 3.354 & $3.040\pm0.040\pm0.150$ \cite{latti2} & \\
$2^3P_0$ & $0^{-+}$ & 3.449 & $3.640\pm0.060\pm0.180$ \cite{latti1} & \\
$1^1G_4$ & $4^{++}$ & 3.858 & $3.650\pm0.060\pm0.180$ \cite{latti3} & \\
         &          &       & $3.608\pm0.110$ \cite{latti4} & \\ 
$1^5G_3$ & $3^{++}$ & 3.895 & $3.670\pm0.050\pm0.180$ \cite{latti2} & \\
$1^3F_2$ & $2^{-+}$ & 3.575 & $3.890\pm0.040\pm0.190$ \cite{latti1} & \\
$1^1I_6$ & $6^{++}$ & 4.552 & $4.339\pm0.261$ \cite{latti4} & \\
\hline		\hline
		\end{tabular}
		\caption{Masses of some low-lying two-gluon glueballs computed with our model (third column) and the parameters of Table~\ref{tabparams}. The quantum numbers of a particular $J^{PC}$ state are summed up in spectroscopic notation in the first column. Our results are compared to lattice QCD calculations (fourth column), and corresponding experimental candidates are suggested in the last column. All the masses are given in GeV.}
		\label{resglu}
\end{table}		

\end{document}